\documentclass[%
 reprint,
 amsmath,amssymb,
 aps,
]{revtex4-2}

\usepackage{color}

\usepackage{hyperref}
\usepackage{graphicx}
\usepackage{dcolumn}
\usepackage{bm}
\usepackage{color}
\usepackage{amsmath}

\begin{document}

\title{Modeling observed gender imbalances in academic citation practices}

\author{Jennifer Stiso$^{1}$}
\author{Kendra Oudyk$^{2}$}
\author{Maxwell A. Bertolero$^1$}
\author{Dale Zhou$^{1}$}
\author{Erin G. Teich$^{1}$}
\author{David M. Lydon-Staley$^{3}$}
\author{Perry Zurn$^{4}$}
\author{Dani S. Bassett$^{2,5,6,7,8,9,10}$}

\affiliation{
$^1$Department of Bioengineering, School of Engineering and Applied Science, University of Pennsylvania, Philadelphia, PA 19104, USA
}
\affiliation{
$^2$Integrated Program in Neuroscience, McGill University, Canada
}
\affiliation{
    $^{3}$ Annenberg School for Communication, University of Pennsylvania, Philadelphia, PA, USA
}
\affiliation{
    $^{4}$ Department of Philosophy, American University, Washington, DC 20016, USA
}
\affiliation{
	$^5$Department of Electrical \& Systems Engineering, School of Engineering \& Applied Science, University of Pennsylvania, Philadelphia, PA 19104, USA
	}
\affiliation{
	$^6$Department of Neurology, Perelman School of Medicine, University of Pennsylvania, Philadelphia, PA 19104, USA
}
\affiliation{
	$^{7}$Department of Psychiatry, Perelman School of Medicine, University of Pennsylvania, Philadelphia, PA 19104, USA
}
\affiliation{
	$^{8}$Department of Physics \& Astronomy, College of Arts \& Sciences, University of Pennsylvania, Philadelphia, PA 19104, USA
}
\affiliation{
	$^{9}$The Santa Fe Institute, Santa Fe, NM 87501, USA
}
\affiliation{
$^{10}$To whom correspondence should be addressed: dsb@seas.upenn.edu
}

\begin{abstract}
In multiple academic disciplines, having a perceived gender of `woman' is associated with a lower than expected rate of citations. In some fields, that disparity is driven primarily by the citations of men and is increasing over time despite increasing diversification of the profession. It is likely that complex social interactions and individual ideologies shape these disparities. Computational models of select factors that reproduce empirical observations can help us understand some of the minimal driving forces behind these complex phenomena and therefore aid in their mitigation. Here, we present a simple agent-based model of citation practices within academia, in which academics generate citations based on three factors: their estimate of the collaborative network of the field (i.e., their understanding of who is in the field and with whom those people collaborate), how they sample that estimate, and how open they are to learning about their field from other academics. We show that increasing homophily---or the tendency of people to interact with others more like themselves---in these three domains is sufficient to reproduce observed biases in citation practices. We find that independent sources of homophily control the static and time-varying aspects of citation bias. More specifically, homophily in sampling an estimate of the field influences total citation rates, and openness to learning from new and unfamiliar authors influences the change in those citations over time. We next model a real-world intervention---the citation diversity statement---which has the potential to influence both of these parameters. We determine a parameterization of our model that matches the citation practices of academics who use the citation diversity statement. This parameterization paired with an openness to learning from many new authors can result in citation practices that are equitable and stable over time. Ultimately, our work underscores the importance of homophily in shaping citation practices and provides evidence that specific actions may mitigate biased citation practices in academia.

\end{abstract}

\maketitle

\section*{Introduction}

Science is ultimately a human endeavor, and must be evaluated in the social contexts of the humans who participate in it. Much of the social context surrounding and within academia contains biases against scientists based on features of their gender (whether self-identified or inferred), as evidenced by gender disparities in hiring and promotion\cite{Casad2021,Shen2013,Charlesworth2019}, mentoring\cite{Moss-Racusin2012}, salary\cite{Shen2013,Moss-Racusin2012,Charlesworth2019}, grants\cite{Shen2013,Beaudry2016}, and awards\cite{Charlesworth2019}. Some work has found that these biases are apparent before any interaction with an individual has occurred, instead arising merely in response to their name\cite{Moss-Racusin2012}. To study how this type of bias influences other outcomes in academia, researchers have used algorithmic approaches to assign names into two bins---woman and man---and then compared outcomes for individuals in each bin. This approach is limited in that it cannot identify outcomes for trans, non-binary, or genderqueer individuals, given that those identities are not reflected in names. However, the name-based approach does have some strengths, in that it can quantify biases that arise from one common dimension of perceived identity. As such, the name-based approach can be an important part of a comprehensive study of gender bias in academia. Using this approach, it has been demonstrated that some scientific outcomes, including citations, vary based on the perceived gender of the name. Specifically, there is a marked undercitation of teams with woman names in neuroscience\cite{Dworkin2020.01.03.894378}, astronomy\cite{Caplar2017}, medicine\cite{Chatterjee2021}, communication\cite{Wang2021}, political science\cite{dion2018gendered} international relations\cite{maliniak2013gender}, and physics\cite{Teich2021}. Undercitation of these teams is problematic as citations are an important currency in academia, influencing promotions, immigration status, whose work is discussed by others, and who is credited as a leader. Citing others is a social practice, forging a network of social connections among scientists who credit, reject, court, and memorialize each other. Individuals with woman names are systematically under-resourced in this currency of academia. 

How can these biases exist and be perpetuated despite more women earning doctorates\cite{Casad2021}? The underlying biased social context of academia can support a variety of explicitly and implicitly biased practices and policies that perpetuate inequity. More specifically, these policies and people's reactions to them interact in complex ways to drive increased scientific interaction with individuals of the same gender. This preference is referred to as gender homophily. The term homophily captures the tendency of individuals to interact more with individuals who are similar to them, and it can be observed between individuals with similar beliefs or similar fields of study, as well as similar identities\cite{McPherson2001,Weare2009}. In academia specifically, gender homophily has been demonstrated in the editorial process \cite{Makin2021} and in co-authorship\cite{Holman2019}. Additionally, academics tend to cite individuals nearby in their co-authorship networks, and these networks influence an individual's inferences about the field as a whole\cite{Lee2019}. Whether these homophilic trends arise through choices or are imposed by institutions\cite{McPherson2001,Greenberg2017}, the trend toward sameness tends to mitigate efforts to rectify disparities by recruitment, and might contribute to why some fields show increasing disparities despite increasing diversity\cite{Baron2016,McGuire2002}. Models that include only gender homophily in directed citation networks can recreate the undercitation of women\cite{Nettasinghe2021}. However, there exists an opportunity to extend such models to include homophily in social networks, which would add to our understanding of how diverse homophilic forces shape citation practices over time. Here, we seek to computationally test the hypothesis that manifestations of gender and social homophily can recreate observed biases in citation practices in academia and identify promising candidate interventions to mitigate existing bias.

Since it is difficult to experimentally test interventions within a social system, computational modelling can be a useful tool for identifying behavioral driving forces and testing potential changes to a system's structure\cite{Momennejad,Verhoeven2020}. Here, we used a computational model to simulate the conditions under which observed citation imbalances could arise. More specifically, we developed an agent-based model\cite{Bonabeau2002} where the sample of artificial academics (\emph{agents}) has a gender of woman or man, and an internal estimate of the authors in their field (Fig. 1A). Agents then go through simulations where they meet and have discussions with other authors, decide whether or not to update their estimate of the field based on their discussions, and then generate citation lists from their estimate of the field. Agents display gender homophily in two ways: (1) in their internal estimates of gender proportions in the field, and (2) in the other agents whose work they discuss and cite. The agents also display scientific topic homphily when deciding which other agents they will learn from. Each agent has 3 parameters that determine the strength of their homophilic tendencies: (1) the proportion of women in their estimate of the field (Fig. 1B); (2) how likely they are to cite or discuss women in their estimate of the field (Fig. 1C); and (3) how similar other authors' estimates of the field have to be to their own in order to learn from them (Fig. 1D). Hence, the full model allows us to determine whether and to what degree these factors could explain observed citation patterns in academia. 

In evaluating the model, we tested for the existence and extent of three outcome measures commonly observed in citations in academia: (1) an undercitation of women authors relative to their prevalence in the field; (2) an increase in this undercitation despite a diversifying field; and (3) a tendency for man citers to drive the undercitation of women authors. We use the field of neuroscience to benchmark our model because of the thorough quantification of citation behaviors in the 5 top journals in the field\cite{Dworkin2020.01.03.894378}. We find that our model produces all 3 observed citation biases. We then test potential interventions that could correct inequities in citation practices. We find that interventions to citation and discussion biases alleviate the undercitation of women, whereas interventions expanding agents' openness lessen the increasing undercitation over time. Lastly, we explore a single empirical intervention---the citation diversity statement---that has the potential to impact the citation gender gap. We parameterize our model to match citation patterns from authors who have used this intervention, and show that achieving equitable citation practices is possible, though it would require collective action. Broadly, our simulation based study provides important evidence for particular drivers of citation imbalances, and quantitative assessments of the efficacy of attitudinal and actionable intervention strategies.

\begin{figure}
	\begin{center}
		\centerline{\includegraphics[width=.45\textwidth]{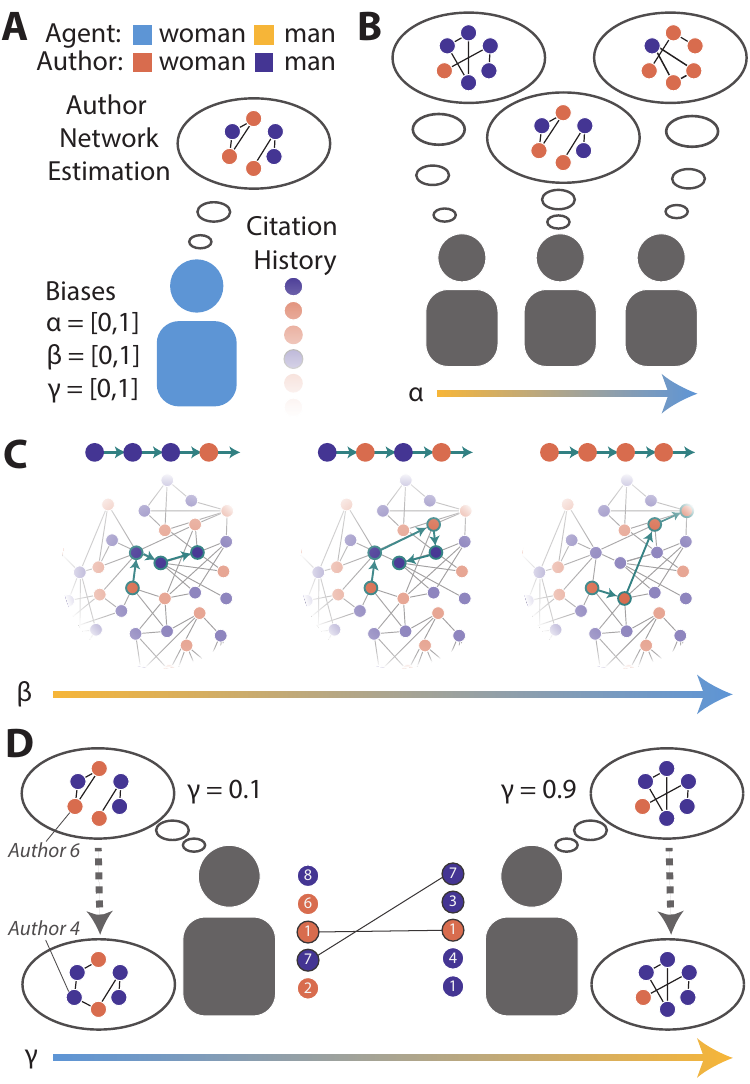}}
		\caption{\textbf{Schematic of model.} \emph{(A)} The properties of each agent. Agent gender is depicted in blue for women and yellow for men, whereas author gender is depicted in orange for women and purple for men. \emph{(B)} A schematic of the effects of $\alpha$, which determines the gender makeup of agent's subjective network estimations. As $\alpha$ increases, network estimates will contain more women. The average value of $\alpha$ is higher for women agents, as indicated by the color gradient on the $\alpha$ line. This parameter operationalizes gender homophily in network estimates, where women agents will tend to think the field contains more women and where men agents will think the field contains more men. \emph{(C)} A schematic of the effects of $\beta$, which determines how the network estimate is sampled. As $\beta$ increases, walks sampling the network will be biased towards women. The average value of $\beta$ is higher for women agents, as indicated by the color gradient on the $\beta$ line. This parameter operationalizes gender homophily in discussions and citations of authors, where women agents will tend to discuss and cite the work of women more and where men will tend to discuss and cite the work of men more. \emph{(D)} A schematic of how $\gamma$ influences learning during simulated meetings. Agents with low $\gamma$ values will have a lower threshold for learning and update their network estimation more often than individuals with high $\gamma$. In this example, the updated post-learning network for one agent replaces Author 6 with Author 4. The average value of $\gamma$ is lower for women agents, as indicated by the color gradient on the $\gamma$ line. This parameter operationalizes scientific homophily in who agents learn from, where agents will only learn new information from other agent's who seem to discuss and cite similar work.}
		\label{fig:schematic}
	\end{center}
\end{figure}

\section*{Methods}
In this work, we developed an agent-based model and simulated the model's dynamics to identify conditions under which biased citation practices might arise. Motivated by direct assessments in the increasingly diversifying field of neuroscience, we hypothesized that three factors would explain observed imbalances: (1) bias towards similar genders in individuals' estimates of who is in the field; (2) bias towards similar genders when discussing or citing other work; and (3) bias towards more similar individuals when deciding from whom to learn about the field. Here, we operationalize inter-individual similarity as having a similar estimate of the field, evidenced by a history of discussing similar authors in our simulations. To appreciate how these factors can produce the observed biases, we first turn to a description of the co-authorship networks that link authors with one another. 

\subsection*{Co-authorship networks}
Investigating our hypotheses requires us to select a representation of the social and collaborative ties within a scientific field. Further, it requires us to select a method of sampling that representation in a manner that can be unbiased or biased towards either gender. Here, we use a graph of co-authorship between authors and leverage tools from graph theory to test our hypotheses. Specifically, we represent the field of neuroscience (or an individual's estimation of it) as a network $C$ of discrete nodes, or authors, connected by edges, or co-authored papers. This co-authorship network is binary in the sense that an edge exists if there is at least one co-authored paper containing both individuals, and is absent otherwise. 

For neuroscience, this network was calculated by extracting data from the Web of Science for research articles and reviews published in five top neuroscience journals (\emph{Nature Neuroscience}, \emph{Neuron}, \emph{Brain}, \emph{Journal of Neuroscience}, and \emph{NeuroImage}) since 1995 \cite{Dworkin2020.01.03.894378}. All unique authors in the dataset are represented by network nodes, and the presence of a co-authored paper between two authors is represented by an edge. Authors with no edges were removed from the dataset. Gender was assigned to authors using the Gender API (gender-api.com), a paid service that supports roughly 800,000 unique first names across 177 countries. We assigned `woman'(`man') to each author if their name had a probability greater than 0.50 of belonging to someone labeled as ‘woman’(‘man’) according to the sources used by Gender API. If the probability was 0.50 exactly, the name would be labeled as `woman'.

After building this representation of the field, we then build each agent's network estimate by sampling sequences of authors from a random walk on the co-authorship network. Random walks start at a given node in the network, and then step randomly to another connected node. These steps are repeated until we have reached a desired walk length. Walks can be biased by changing the probability with which we step to a new node, based on some property of that new node. Here, probabilities will be biased based on the gender of authors.

\subsection*{Author definitions}
To test our hypotheses, we began by simulating an initial population of 200 agents. In this initial sample, the proportion of women agents was 36\%, equivalent to the proportion of women authors in neuroscience in the year 1995. Here, we use the term \emph{agent} to refer to a simulated author who participated in meetings, generated citations, and performed other actions. We differentiate agents from \emph{authors} in the co-authorship network that the agents can cite. Each agent has 3 free parameters that determine (1) the proportions of each gender in their network estimate of the field ($\alpha$); (2) the likelihood that the agent will discuss the work of each gender in meetings ($\beta$); and (3) the amount of overlap in the discussed and cited authors needed for this agent to learn new authors from one another ($\gamma$). 

Each agent is also initialized with a distinct citation history and with a distinct estimation of (i.e., sample from) the co-authorship network in neuroscience. Estimations are biased towards a given gender with the parameter $\alpha$. The citation history is a running history of the authors that this author has previously cited or discussed, which is biased towards a given gender with the parameter $\beta$. The parameters $\alpha$ and $\beta$ are drawn from different distributions depending on the gender of the agent. In alignment with literature showing homophily in co-author and citation networks\cite{Holman2019,Lee2019}, women are given higher values of $\alpha$ and $\beta$ compared to men, leading them to know and cite more women authors. After each meeting or citation, this history is updated. It can then be used by other agents to determine if they are willing to learn from this agent. The level of similarity needed for a successful meeting is determined by the parameter $\gamma$. Since minorities have been shown to have more innovative and interdisciplinary research programs\cite{Hofstra2020}, this parameter is lower for women agents than men agents.

\subsection*{Simulated meetings}
In simulations, each agent will meet with all other agents. If an agent determines that the other agent is similar enough to them, they will learn from a discussion with that agent\cite{Ding2011}. Discussions consist of one agent generating a list of 70 authors from their network estimate, which updates their own citation history. This length was selected because it is comparable to the average number of citations used for an empirical or review article in neuroscience. Since authors' citations are influenced by other researchers\cite{Purwitasari2020}, the other agent will then incorporate some of the discussed authors into their network estimate, and forget some of the authors in their network estimate that are rarely discussed. The number of authors learned and forgotten are equal, ensuring that results are not driven by changes in the size of network estimates. The number of authors learned with each meeting is determined by the parameter $\zeta$. We refer to this process as the second agent learning from the first. 

Meetings will have a temporal organization. Each agent will have 10 meetings per simulated year. After each year, every author generates a paper, with a reference list of length 70 and updates their citation history. These citations are then used to calculate the percentage of overcitation at this time point. Lastly, new authors are added to the field in a manner that is consistent with the increasing diversity of neuroscience (and most STEM fields); specifically, in our simulations we add about 1 new woman agent into the field each year. This rate was selected to recreate a magnitude of gender diversity increase similar to that which has been reported for neuroscience (0.61\% per year) \cite{Dworkin2020.01.03.894378}. The process was repeated over 23 years, giving a final population of 256 agents that are 50\% women. 

\subsection*{Bias ($\alpha$) in the estimation of the author network}
Agents' network estimations ($\hat{C}$) are determined by a biased diffusive random walk of length 500 steps on the true co-authorship network in neuroscience ($C$). To determine the step size of the diffusive random walk, we used an exponential (Pareto) distribution which results in many small steps and a few large leaps to disconnected regions. Unlike a random walk, this method allows each author to have an estimate of the field that contains a few distinct groups of co-authors. The distribution is parameterized by two free parameters, $\mu$ and $d$, where $\mu$ determines the distribution's rate of decay, and $d$ determines the maximum step size. For our author population, $\mu$ was set to 3 and $d$ was set to 3. These parameters were chosen to ensure that authors had multiple---but few---connected components in their final estimates of the field\cite{Robinson2014}.

Diffusive walks were also biased towards a given gender depending on the parameter $\alpha$, which ranges from 0 to 1. Specifically, $\alpha$ determined the probability distribution of possible nodes for the next step of the walk such that:

\begin{equation*}
	P(l) = 
	\begin{Bmatrix}
		\frac{\alpha}{z} & l_{gen} = \mathrm{woman}\\
		\frac{1 - \alpha}{z} & l_{gen} = \mathrm{man}
	\end{Bmatrix}, 
\end{equation*}

\noindent where $l$ is a possible author connected to the current author, whose gender ($l_{gen}$) is $woman$ or $man$. The probability of selecting author $l$ is given by $P(l)$, and $z$ is a normalization constant used to ensure that the probabilities of selecting each connected author sum to 1. With this definition, if $\alpha = 0.5$, the probabilities for man and woman nodes would all be equal, and the successive nodes in the walk would be selected randomly from the set of possible nodes. If $\alpha = 1$ ($0$), walks would always choose women (men) when possible. If an author is connected only to authors of one gender, the next step will be drawn randomly from these authors regardless of the values of $\alpha$.

\subsection*{Bias ($\beta$) in which authors are discussed}
Discussions of other authors occur both explicitly and implicitly. Explicit discussion of other authors occurs in meetings, conferences, emails, and Twitter, for example; implicit discussions of authors occurs in the citations or reference lists provided in written material. Here in our model, such discussions are determined by a biased random walk on the agent's estimate of the network. Since these walks are not diffusive, all step sizes are equal to 1. The parameter $\beta$ determines the probability of selecting women or men as the next step in the walk where:

\begin{equation*}
P(l) = 
\begin{Bmatrix}
\frac{\beta}{z} & l_{gen} = \mathrm{woman} \\
\frac{1 - \beta}{z} & l_{gen} = \mathrm{man}
\end{Bmatrix} , 
\end{equation*}
\noindent where $l$ is a possible author connected to the current author, with a gender ($l_{gen}$) of $woman$ or $man$. The probability of selecting author $l$ is given by $P(l)$, and $z$ is a normalization constant to assure that the probabilities of selecting each connected author sum to 1. With this definition, if $\beta = 0.5$, the probabilities for man and woman nodes would all be equal, and the successive nodes in the walk would be selected randomly from the set of possible nodes. If $\beta = 1$ ($0$), walks would always choose women (men) when possible. If an author is connected only to authors of one gender, the next step will be drawn randomly from these authors regardless of the values of $\beta$.

\subsection*{Bias ($\gamma$) in which authors are learned from}
Agents have the opportunity to meet with other agents, but will only learn new authors from those who have network estimates that seem similar to their own. The threshold---that defines how similar other agents have to be to allow learning to occur---is determined by the parameter $\gamma$. The goal of $\gamma$ is to allow us to simulate individuals' increased openness to those who study different topics and have different social networks\cite{CASSANDRO2010}. In our model, those beliefs are agents' internal estimates of the co-authorship network. However, outside of these simulations, those true beliefs are not accessible to any individual other than the one who holds them. Therefore, instead of comparing network estimates, we compare the citation history of two agents, or their records of discussed or cited authors. In a meeting between two agents, if the percent of authors shared across their citation histories is greater than the first agent's $\gamma$, then that first agent will learn from the second, and update their network estimation with new authors from the second agent's discussion. The same process applies to the second agent, when determining whether to learn from the first agent. Minorities in science tend to produce research that is more interdisciplinary, consistent with an increased openness to learning from scientists with diverse backgrounds\cite{Hofstra2020}. Therefore, we set $\gamma$ to be lower in women than in men.

\subsection*{Learning new authors ($\zeta$)}
The $\zeta$ parameter influences how agents learn new authors during meetings. During a successful meeting, an agent will learn or update their network estimation $\hat{C}$, based on a sequence of discussed authors from another agent. Here, we assume that the receiving member of the discussion is attempting to remember not only the listed authors, but the underlying relationships between them as defined by the network $A$. Additionally, we assume that the authors that are remembered will be the ones that appear to be central to the other agent's network estimate. Empirically, individuals learn network structure from sequences with different levels of fidelity. This fidelity can be modeled via updates to an internal estimation of the underlying space of possible co-authorships between academics, where updates are defined by a decaying memory function across previously experienced citations. In this model, low levels of fidelity would result in highly inaccurate knowledge of the relationships between items; high levels of fidelity would result in learning the true relationships between authors; and middling levels of fidelity would result in selectively higher memory for specific relationships, such as clusters\cite{Lynn2020}. More specifically, given an observed sequence of authors $a_1, ... ,a_{t}$, and given a value of $\zeta$, this model predicts agents' internal estimates of transition probabilities between authors $a_i$ and $a_j$ as $\hat{A}_{ij}(t)$. The estimated entries of $\hat{A}$ could be obtained with the following equation: $\hat{A} = (1 - e^{-\zeta})A(I - e^{-\zeta}A)^{-1}$, where $A$ is the true co-authorship structure between authors. This analytic prediction reflects the estimated co-authorship space for an agent that viewed an infinite random walk on the co-author network, and does not take into account the statistics of the particular sequence observed by a given agent. 

We now need to select some authors to remember, and we choose those that are central to the estimate $\hat{A}$. We define central authors as those who have many strong connections to other authors in $\hat{A}$. Specifically, all agents will add any authors with edges whose weights are greater than 0.1 in $\hat{A}$ to their network estimate $\hat{C}$. This way, the number of authors learned will vary with $\zeta$ in a way that reflects individual differences in sequential learning as demonstrated through behavioral experiments. In order to prevent learned authors from being disproportionately isolated from others in the network, empirical co-authorships between newly added authors and pre-existing authors are added once the most strongly connected authors are selected. 

\subsection*{Quantifying overcitation}
To evaluate the citation behavior of our sample of agents, we calculate the overcitation percentage for each gender within each agent's reference list. The overcitation percentage for women is defined as $overcitation_w = \frac{obs_w - exp_w}{exp_w}$, where $obs_w$ is the observed proportion of women cited in this reference list and where $exp_w$ is the expected proportion of women, or the percentage of women authors in neuroscience in a given year.

\subsection*{Statistical analyses}
Throughout the manuscript, we assess the statistical significance of two features of our simulations: the overcitation of authors of a specific gender, and the change in overcitation across meetings. Significant overcitation is detected using a $t$-test, and significant changes in overcitation across meetings are detected with a linear model in which overcitation is the dependent variable and time is the independent variable.

\subsection*{Citation diversity statement}
In order to validate the results of our agent-based model with real-world observations, we sought to identify values of $\beta$ that would reflect authors who adopt a citation diversity statement into their work. To do this, we identified papers that included citation diversity statements in the currently published literature and then calculated their overcitation rates for papers authored by women and/or men. We gathered these real-world data by searching for papers that cited the paper introducing citation diversity statements\cite{Dworkin2020.01.03.894378} and/or a tool used to create such a statement\cite{zhou_dale_2020_3672110}. This search was done manually using using Google Scholar and OpenCitations.net. We identified 141 unique citing papers, 89 of which contained citation diversity statements. We also developed an automatic method for collecting such papers using CrossRef's API, but we found that the manual search resulted in a more complete list of citing papers, particularly for papers published very recently.

After gathering the papers, we extracted the reported citation rates of 4 different categories of papers from those papers' citation diversity statements: papers with woman first and last  authors (ww); man first, woman last (mw); woman first, man last (wm); and man first and last (mm). Then, we defined the overcitation of each author-gender category as $overcitation_{ww} = \frac{rep_{ww} - exp_{ww}}{exp_{ww}}$, where $rep_{ww}$ is the reported proportion of cited papers with woman first and last authors, and $exp_{ww}$ is the expected proportion of such papers. For the latter, we used the benchmark expected citation rates reported in the paper introducing the citation diversity statement \citep{Dworkin2020.01.03.894378}, which were 6.7\% ww, 9.4\% mw, 25.3\% wm, and 58.6\% mm.

\subsection*{Data and Code}
Code is available at github.com/jastiso/modeling\_citations. Code and data for calculating citation rates in papers using the citation diversity statement is available at github.com/koudyk/cleanBibImpact.

\section*{Results}
In this work, we sought to develop a model that would recreate the observed imbalance in citation practices. Specifically, we sought to recreate three observations: (1) the undercitation of women and overcitation of men; (2) the bias being driven by the majority gender (men); and (3) the increase in bias over time. We hypothesized that in an increasingly diversifying field, three factors would recreate all observations: (1) an individual's estimates of who is in the field tend to be skewed towards the agent's own gender; (2) when discussing or citing other authors, an individual preferentially mentions authors of the individual's own gender; and (3) an individual is more likely to update their own beliefs when receiving information from others who are similar to themselves.

To simulate the observed biases in citation practices in neuroscience, we chose initial values for all of our parameters based on findings from neuroscience\cite{Dworkin2020.01.03.894378}. This paper demonstrates that authors tend to overcite men and undercite women. Women tend to overrepresent women in their co-authored papers by 0.8\% $\pm$ 0.1\%, while men tend to overrepresent men in their co-authored papers by 4.4\% $\pm$ 0.1\%. We select the parameter $\alpha$ to have deviations from 0.5 (no bias) that reflect these findings. Specifically, $\alpha$ is centered at 0.51 (1\% bias) for women and 0.45 (5\% bias) for men, with a standard deviation of 0.01 (Fig. \ref{fig:static}A, left). 

Women-led teams also tend to overcite women by 9.5\% $\pm $5.5\%. Men tend to overcite men by 5.9\% $\pm$ 0.4\%, after accounting for the overrepresentation of men in their co-authorship networks. We therefore set $\beta$, the sampling parameter, to be centered at 0.6 for women (10\% bias) and 0.44 (6\% bias) for men, with a standard deviation of 0.1 (Fig. \ref{fig:static}A, middle). The value 0.1 was the largest standard deviation that reliably gave values between 0 and 1. 

Minorities in science tend have work that is more interdisciplinary\cite{Hofstra2020}. Therefore, we set the value of $\gamma$ to be lower for women than for men. The value of $\gamma$ was set to a strict threshold in order to produce selectivity in which agents meet (Fig. S2). Women's distributions were centered at 0.04 and men's were centered at 0.06, with a standard deviation of 0.005 (Fig. \ref{fig:static}A, right). The learning rate $\zeta$ was centered at $0.1$, in line with work quantifying this parameter in statistical learning tasks in humans\cite{Lynn2020}. In the supplement, we show that results are robust for different selections of $\gamma$ and $\zeta$ (Fig. S3). We then simulated 23 years with 10 meetings per year between all pairs of agents and collected reference lists from each agent after every year. 

\subsection*{Citation imbalance driven by men}
We first quantified the overcitation of each gender by our population of agents. After each round of meetings, we calculate the overcitation of women and men for each author. We then average this overcitation across all time points, leaving us with one overcitation value for each author. Our sample of agents tends to cite women 25.8\% $\pm$ 2.66\% less than expected, and tends to cite men 21.0\% $\pm$ 2.11\% more than expected (Fig. \ref{fig:static}B). We find that the citation of women is statistically below zero (one-sample $t$-test, $t(256) = -15.9$, $p = 5.13 \times 10^{-40}$). This analysis demonstrates that our model recreated empirically observed disparities in the citation of women and men scholars.

Next, we identified which gender was driving the imbalance in citation. Here, we find that man agents have greater citation imbalance. Men cite men 30.5\% $\pm$ 1.81\% more than expected, and men cite women 38.5\% $\pm$ 2.25\% less than expected (one-sample $t$-test $t(256) = -22.5, p = 2.1\times10^{-46}$). Women cite women 9.27\% $\pm$ 2.79\% less than expected, and women cite men 8.60\% $\pm$ 2.23\% more than expected (one-sample $t$-test $t(256) = -5.51, p = 1.91\times10^{-7}$). While this undercitation of women authors by women agents is small, it is inconsistent with data demonstrating that women authors tend to overcite women. Additionally, this analysis demonstrates that our model replicates the observation that men drive citation imbalances.

\subsection*{Citation imbalance worsening over time}

\begin{figure}
	\begin{center}
		\centerline{\includegraphics[width=.45\textwidth]{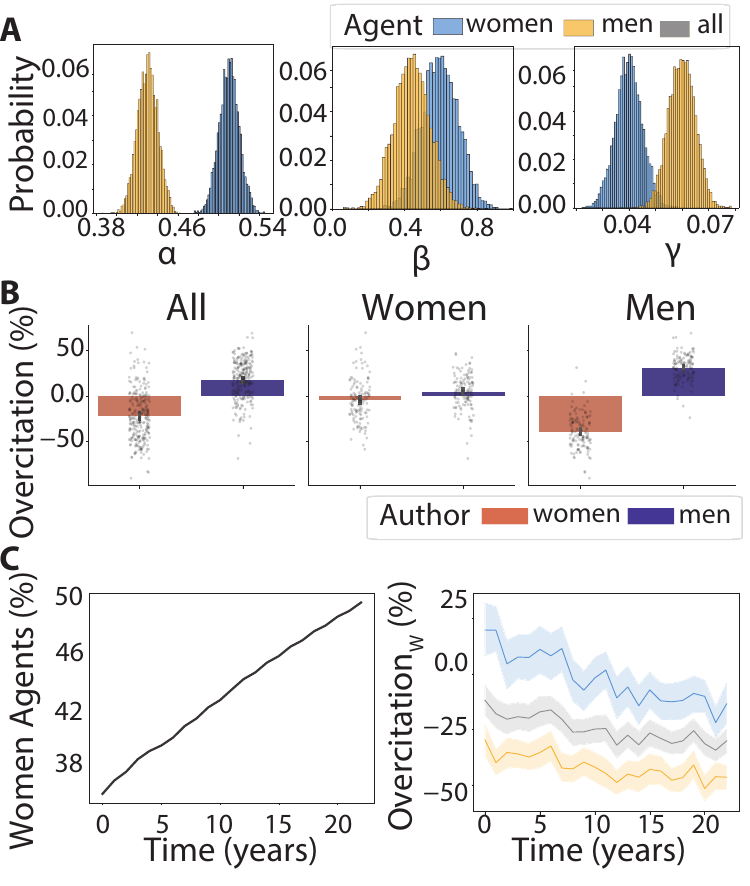}}
		\caption{\textbf{Replicating citation bias.} \emph{(A)} The distribution of parameter values used to simulate citation practices. The parameters from left to right are $\alpha$, $\beta$, and $\gamma$. Distributions of parameter values for women agents are shown in blue and for men agents are shown in yellow. \emph{(B)} The overcitation of authors of each gender from all agents (left), women agents (center), and men agents (right). Woman authors are shown in orange and man authors are shown in purple. \emph{(C)} On the left, we show the change in the percent of women agents over time (left). On the right, we show the overcitation of women over time for each agent gender (yellow, blue) and for all agents (grey).}
		\label{fig:static}
	\end{center}
\end{figure}

Our simulations incorporate the observed increase in women authors in neuroscience. This means that at each successive time point, agents will have the opportunity to learn from more women, who are in turn more likely to discuss the work of women. Additionally, the expected proportion of women cited increases over the 20 year period (Fig. \ref{fig:static}C). To understand how these two forces affect overcitation, we calculate overcitation using the same method described above, but do not average over points in time. We find that over time, the overcitation of women becomes more negative. This downward trend is seen in both women and men agents, but is more pronounced for women agents, who start with positive overcitation values (linear model $overcitation_{w,w} \sim time: \beta=-1.55\times10^{-2}, p = 7.51\time10^{-13}$, $overcitation_{w,m} \sim time: \beta = -9.75\times10^{-3}, p = 1.11\times10^{-9}$). This indicates the woman agents cite women 1.6\% less after each year, and man agents cite women 0.98\% less after each year. This trend persists in agent populations that are majority women (see Fig. S2). Therefore, our model recapitulates the increasing citation disparity observed in some diversifying fields.

\subsection*{Interventions to biased discussions and citations}

\begin{figure*}[!ht]
	\begin{center}
		\centerline{\includegraphics[width=\textwidth]{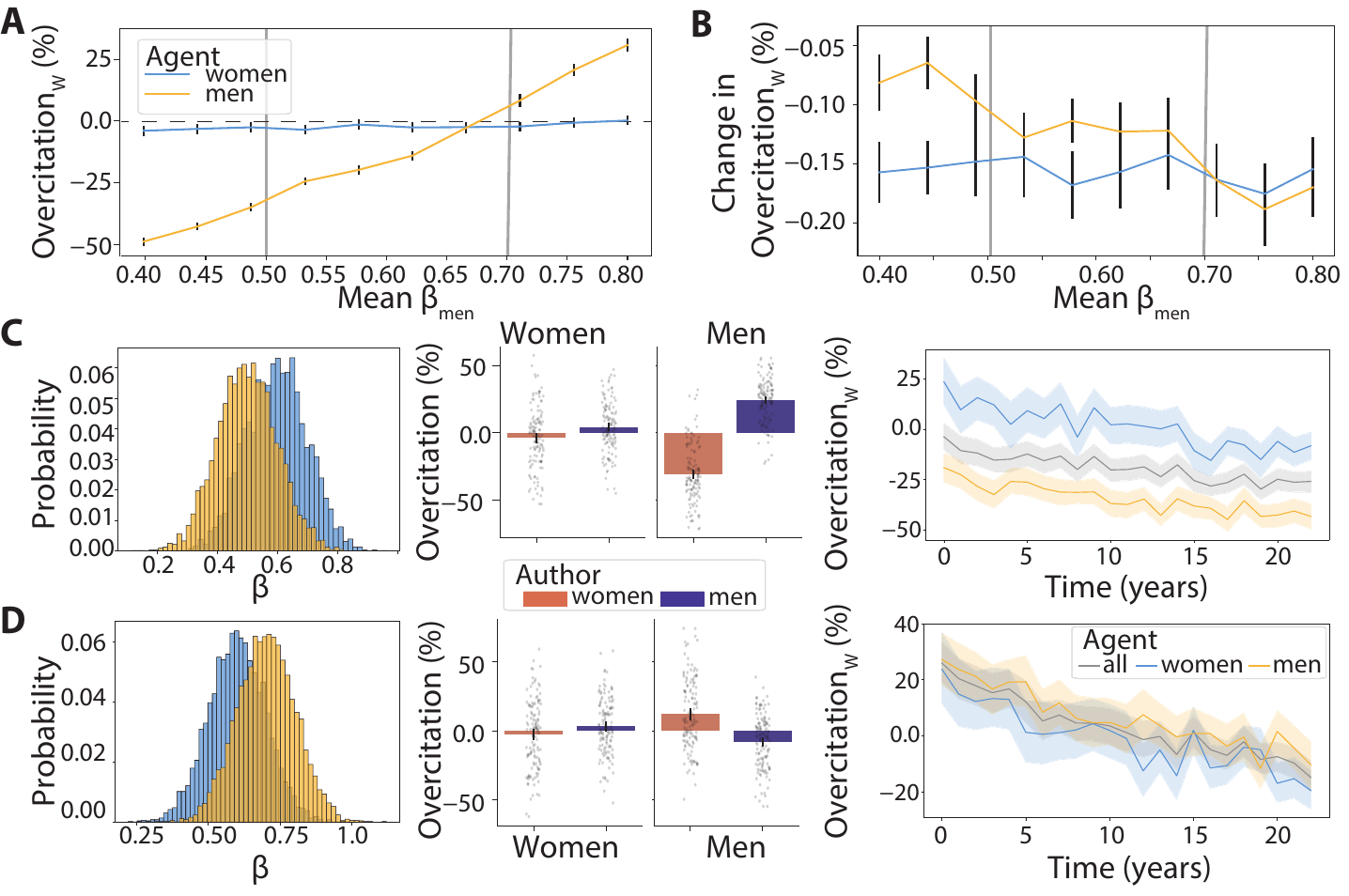}}
		\caption{\textbf{Interventions to network sampling ($\beta$).} \emph{(A)} The total overcitation of women for women (blue) and men (yellow) agents as the mean value of men's $\beta$ changes. Grey lines indicate parameter values that are shown in panels \emph{(C)} and \emph{(D)}. \emph{(B)} The same information as in panel \emph{(A)}, but visualizing the slope of overcitation of women rather than the total overcitation. \emph{(C)} The total (center) and time varying (right) overcitation when men's $\beta$ values are centered on $0.5$ (left). \emph{(D)} The total (center) and time varying (right) overcitation when men's $\beta$ values are centered on $0.7$ (left).}
		\label{fig:int1}
	\end{center}
\end{figure*}

It is possible that changes to the behavior of agents could lead to sustained, equitable citation practices within our sample. We define equitable citation practices as those that are citing at rates reflective of gender ratios within the field or with some overcitation of women to account for the history of inequality. In order to understand how agents' behavior can be altered to achieve equitable citations, we simulate interventions that change the underlying distributions of some parameters. Here, we wish to focus on parameters that could be changed by real world policies around citation behavior. Therefore, we test interventions to the parameters $\beta$ and $\gamma$, but do not test interventions to the parameter $\alpha$, since it is not feasible to reinstantiate all scientists with less biased internal representations of the field. We also focus our interventions on man agents, since their citation practices are driving the disparity we are studying\cite{Dworkin2020.01.03.894378,wang2021gendered,Teich2021,maliniak2013gender,dion2018gendered,mitchell2013gendered}. 

We first test changes in the parameter $\beta$, which determines how agents sample their internal representation of the network. Specifically, we select 10 $\beta$ values to center the parameter distributions for men ranging from 0.4 to 0.8. For each parameter, we then calculate (1) the average overcitation of women and (2) the average slope of overcitation over time. These two calculations allow us to assess whether each intervention is addressing citation disparities, their worsening over time, or both. Overall, we find that increasing the $\beta$ value of men increases the overcitation of women (Fig. \ref{fig:int1}A), but does not reduce, and actually strengthens, the citation deficit over time (Fig. \ref{fig:int1}B). This stronger citation deficit could be due to a floor effect, where agents with the lowest citation of women overall have little room to decline. We also note that a $\beta$ value of around 0.5, which indicates no bias in citations or discussions, is not enough to ensure equitable citations (Fig. \ref{fig:int1}C, $t$-test $t(256) = -23.5, p = 2.57\times10^{-48}$). The smallest $\beta$ where equitable citations in men arise is around 0.65, which corresponds to a preference for citing and discussing women (Fig. \ref{fig:int1}D, $t$-test $t(256) = 2.89, p = 4.57\times10^{-3}$). This simulation demonstrates that increasing the likelihood of discussing and citing women's scholarship above their representation in the field leads to more equitable citation practices, but does not reverse temporal trends.

\subsection*{Interventions to biased meeting preferences}
\begin{figure}
	\begin{center}
		\centerline{\includegraphics[width=.45\textwidth]{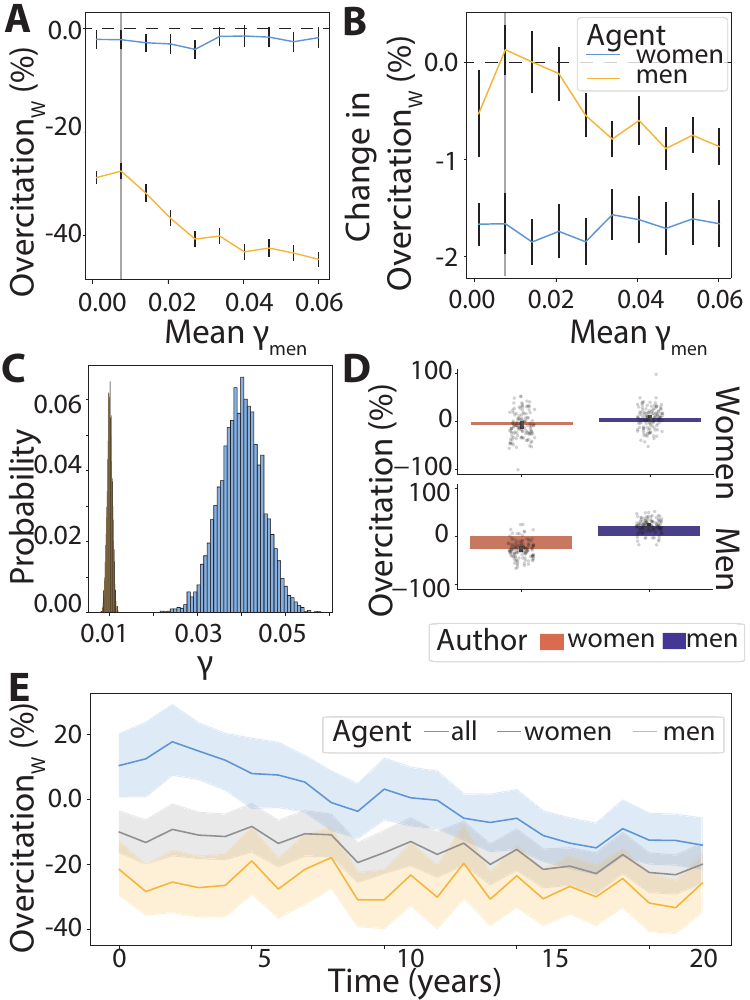}}
		\caption{\textbf{Interventions to meeting tolerance ($\gamma$)}. \emph{(A)} The total overcitation of women for women (blue) and men (yellow) agents as the mean value of men's $\gamma$ changes. Grey line indicates the parameter value that is shown in panels \emph{(C)}-\emph{(E)}. \emph{(B)} The same information as in panel \emph{(A)}, but visualizing the slope of overcitation of women rather than the total overcitation. \emph{(C)} The parameter distribution that shows the most equitable citation practices, where $\gamma$ for men agents is set to $0.01$. \emph{(D)} The total overcitation of woman (orange) and man (purple) authors by women (top) and men (bottom) agents. \emph{(E)} The time varying overcitation of women for women and men agents. Overcitation values averaged across all agents are shown in grey.}
		\label{fig:int2}
	\end{center}
\end{figure}

We next simulated the effects of intervention to the meeting preferences of men. Here, we select 10 values for $\gamma$ that range between 0.001 and 0.06 (the original value used). The value of 0.001 effectively means that agents will learn from all other agents, even if they share no authors in their citation histories. Broadly, we demonstrate that increasing the selectivity of agents that men learn from decreases an already negative overcitation of women (Fig. \ref{fig:int2}A), and leads to sharper decreases in the overcitation of women over time (Fig. \ref{fig:int2}B). We next examine a single value of $\gamma$ that produced the most equitable citation practices (Fig. \ref{fig:int2}C, $0.01$). We find that for $\gamma$ values of $0.01$, man agents still significantly undercite women (Fig. \ref{fig:int2}A and D; $t$-test, $t(256) = -17.9, p = 1.30\times10^{-36}$). However, we find that man agents also do not significantly decrease their overcitation of women over time (Fig. \ref{fig:int2}E; linear model $overcitation_{w,m} \sim time$ $\beta = -2.38\times10^{-3}, p = 0.09$). This simulation demonstrates a marked effect of increasing agents' openness to learning on the temporal properties of citation bias, but shows that changing this behavior alone is not enough to establish equitable citation practices in our agent population. 

\subsection*{Simulation of the citation diversity statement}
Combined, our simulation results suggest that interventions which selectively target citation practices or meeting preferences alone cannot quickly achieve equitable citations. One could instead design a single intervention that spurred change in both dimensions. A key example is the citation diversity statement\cite{Zurn2020}, which is a real-world intervention in which authors typically quantify their citation practices in each reference list they generate. This process gives agents an opportunity to update their bibliographies to select citations that reflect the base rates of women and men in their field before publication, and therefore produce more bibliographies that are at or near equity. While citation diversity statements do not necessitate an increased openness to new authors, thoughtful engagement with the quantification of one's citation practices could lead to updated network estimates in two ways: individuals might be more likely to seek out and remember the work of women at conferences to help generate equitable citations; and individuals might learn about and remember the work of new women authors while refining their citation lists.

\begin{figure}[!t]
	\begin{center}
		\centerline{\includegraphics[width=.43\textwidth]{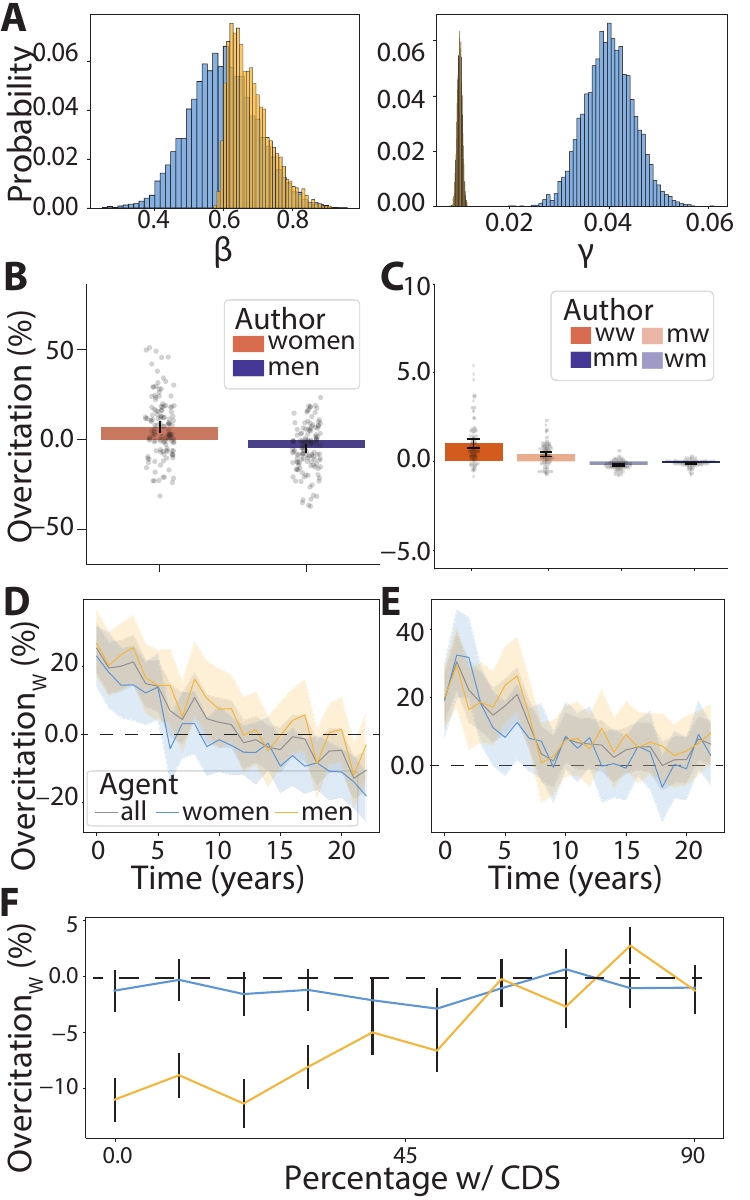}}
		\caption{\textbf{Modeling the citation diversity statement.} \emph{(A)} The distribution of $\beta$ and $\gamma$ values chosen to model the introduction of the citation diversity statement to man agents. The distribution of each parameter is shown in blue for women agents and in yellow for man agents. \emph{(B)} The total overcitation of women authors (orange) and man authors (purple) by men agents in our simulation. \emph{(C)} The observed overcitation of authors using the citation diversity statement as of November 2021. Here, citations are split into first/last author teams. Woman/woman teams are shown in dark orange; man/woman teams are shown in light orange; man/man teams are shown in dark purple; and woman/man teams are shown in light purple. \emph{(D)} The time varying overcitation of women by women agents (blue), men agents (yellow), and all agents (grey) when agents have 10 meetings per year. \emph{(E)} The time varying overcitation of women by women agents (blue), men agents (yellow), and all agents (grey) when agents have 20 meetings per year. \emph{(F)} The overcitation of women as the proportion of inequitable citers using the citation diversity statement (CDS) increases.}
		\label{fig:int3}
	\end{center}
\end{figure}

Accordingly, we wished to simulate interventions to $\beta$ and $\gamma$ that could achieve sustainably equitable citation practices. In choosing those parameters, we were motivated by real-world behaviors that were encouraged by the citation diversity statement. Specifically, we wished to select parameters where nearly all agents were biased towards citing women from their network, but where most agents had small biases and few agents had large biases. For all authors, we construct a new $\beta$ distribution centered at $0.6$, which reflects a slight bias towards women when sampling the agent's network. The distribution is a skewed normal distribution with a skewness value of $10$, a mean of $0.6$, and a standard deviation of $0.1$, such that more than half of the agents have $\beta$ values larger than $0.6$ (Fig. \ref{fig:int3}A, left). We also wished to model an increased openness to engaging with women's scholarship. Therefore, the $\gamma$ value was selected to be centered on $0.01$, which demonstrated the most adaptability to increasing diversity in our previous analysis (Fig. \ref{fig:int3}A, right). Using these parameters, we find that men agents significantly overcite women by 1.56\% (Fig. \ref{fig:int3}B; $t$-test $t(256) = 2.28, p = 0.02$). We can validate our predicted static citations using real data from authors who have used citation diversity statements until November 2021 (Fig. \ref{fig:int3}C, see methods). In empirical data, we calculate citations in 4 groups based on the first and last author of papers: woman first and last (ww); man first, woman last (mw); woman first, man last (wm); and man first and last (mm). We find that our model produces slightly optimistic results compared to the empirical citations of woman and man last authored teams. Specifically, our model shows 5.36\% $\pm$ 3.72\% overcitation of women and 3.04\% $\pm$ 2.90\% undercitation of men, while the empirical data shows 1.14\% $\pm$ 0.14\% overcitation of all woman teams and 0.12\% $\pm$ 0.02\% undercitation of all man teams. 

We find that with our parameterization, man agents still show decreasing citation of women over time (Fig. \ref{fig:int3}D, $overcitation_{w,m} \sim time$ $\beta = 1.39\times10^{-2}, p = 5.86\times10^{-9}$). This observation indicates that more effort is needed to maintain overcitation of women in a diversifying field than to maintain undercitation. We hypothesized that increasing the number of meetings per year might give agents more opportunities to learn, and lead to more equitable citations over time. Therefore, we reran our analysis with 20, rather than 10, meetings per year. While we still observe a decrease in the citation rate of women over time, this value plateaus at just above $0\%$ overcitation (Fig. \ref{fig:int3}E, linear model $overcitation_{w,m} \sim time$ $\beta = 8.58\times10^{-3}, p = 1.34\times10^{-4}$, $overcitation_{t=20} = 9.88$). This trend indicates that increasing the number of meetings is another possible method for keeping citation distributions equitable over time. Additionally, it shows that maintaining stable and equitable citations of women over time requires agents to be more open than they would need to be to simply stabilize inequitable practices. 

Lastly, we sought to identify the minimum proportion of man agents using the CDS that would be needed to reach equitable citations. We find that a large majority, around 80\% of man agents need to adopt the citation diversity statement in order to consistently observe equitable citations (Fig. \ref{fig:int3}F). To summarize our findings in this section, we note that we find a parameterization of one candidate intervention---the citation diversity statement---that matches empirical data. Using this parameterization, we find that equitable citation practices can be achieved, if this intervention is adopted by nearly all of the population.

\section*{Discussion}
Here, we use a simple agent-based model to recreate observed gender disparities in academic citations. Agents display two types of homophily---gender and scientific---controlled by three parameters. We show that in a diversifying field, the maintenance of these two homophilic tendencies is enough to recreate both the gender disparities in citations observed in academia and their worsening over time\cite{Dworkin2020.01.03.894378,Wang2019}. Importantly, we show that even if man citers tend to cite from their own field estimates equitably, with no bias towards their own gender, the same pattern of underciting women emerges. This finding supports the idea that individuals and institutional policies should strive to actively cite, and remove barriers to the citation of, women authors outside of their own networks, rather than cultivate indifference to gender information\cite{Cheryan2020, Lundine2019}.

We also test a few interventions that change the degree of gender and scientific homophily that determines which authors the agents discuss and who they learn from. We find that reducing gender homophily in how agents sample their estimate of the field will increase the citation of women, but does not alleviate the worsening citation rates over time. We find that more learning from individuals outside of the agent's own field estimate is an effective method for keeping pace with changes in demographics, and that more (rather than fewer) meetings are needed to maintain higher (rather than lower) levels of citation for women. Therefore, our model suggests that policies which promote both changes to individual citation practices and changes to how often the work of gender minorities is promoted would mitigate biased citation practices, albeit in different ways\cite{Llorens2021}. Lastly, we simulate a real world intervention with potential to affect both of these biases: the citation diversity statement. We find that intuitive parameter choices match empirical data on the actual effects of citation diversity statements in STEM papers, and we posit how its use might affect changes in citation practices over time. We then ask what portion of man agents would need to adopt this intervention in order to see equitable practices, and we find that a large majority---close to 80\%---are needed. Ultimately, our work provides important theoretical contributions to our understanding of citation practices in academia, and presents actionable hypotheses regarding how best to mitigate them.

\subsection*{Gendered differences in homophily and interventions}
In this work, we design simulations where both woman and man agents demonstrate gender and topic homophily\cite{McPherson2001}. Similar to intersections between genders in diverse social and professional situations\cite{Laniado2016,Kleinbaum2011}, agents tend to estimate that the field is made up of more authors of their own gender than it actually is, and tend to more frequently discuss and cite authors of their own gender within their estimate of the field. Additionally, all agents tend to meet with individuals who share their view of the field, which---although not explicitly gendered---can serve to reinforce existing gender disparities. While all agents show these homophilic tendencies, research suggests that the motivations and drivers of these tendencies are diverse and might not be the same for women and men genders. 

Gender homophily can be induced by institutions or can be a choice of individuals\cite{Greenberg2017,Kleinbaum2011}. Within choice homophily, different genders can also have different motivations for and outcomes from their choices. For example, women or other minorities might employ so-called \emph{activism homophily} in which they increase support to individuals who face systemic barriers similar to those they themselves face\cite{Greenberg2017}. This effect, unlike men's homophily, leads to women receiving more support in areas where they are least represented. This type of homophily ultimately can serve to create more justice and equity in a variety of different spaces including venture funding\cite{Greenberg2017}. It is possible that activism homophily might underlie the observed citation practices of women academics, although to our knowledge that hypothesis has not been explicitly tested. While our model does not distinguish between different drivers of homophily in academia, mapping out such distinctions is a promising direction for future research in this scholarly space. Notably, our model recreates the observed undercitation of women authors by man authors, but does not show overcitation of women authors by women authors. Accounting for gender specific forms of homophily, such as activism homophily, might produce an extended model that better tracks the empirical data. Further investigation of these motivations underlying homophily could also provide insight on which policies might most effectively mitigate bias. 

The fact that men's and women's homophilic tendencies can have opposing forces on equity partially motivated our decision to target men rather than women in interventions. Our decision is also supported by the empirical observation that men's citation practices are driving most of the observed gender imbalance in citations\cite{Dworkin2020.01.03.894378,Wang2019,dion2018gendered,mitchell2013gendered,maliniak2013gender} (see also \cite{Bertolero2020} for evidence that the White authors are driving more of the observed racial imbalance in citations than groups of color). However, the drivers of citation imbalance in any gender are likely multifactorial and might interact with one another in nonlinear ways. In general, we see great value in academics of any gender choosing to critically evaluate their own citation practices and consider what practices can best mitigate their own biases.

\subsection*{Implications for policy}
Our simulated interventions suggest some promising strategies for developing policies to alleviate biased practices in citations. First, our interventions show that the total undercitation of women---and the increasing undercitation of women over time---could arise from separable mechanisms. In our model, the increasing undercitation of women is a result of rigidity in citation practices despite a diversifying field, as observed empirically \cite{Dworkin2020.01.03.894378}. In line with previous modeling work\cite{Nettasinghe2021}, this mechanism suggests that increasing the representation of women in academia will not address citation inequities, and underscores the need for further policy changes beyond increasing representation, especially at junior levels\cite{Kang2019}. 

Additionally, we demonstrate that explicitly forcing agents to sample their own estimate of the field in an unbiased manner is not enough to alleviate biases in citation practices. Since agents still have biased estimates, and because the field is diversifying at a faster rate than agents are updating their estimates of the field, explicit oversampling of women is needed to overcome total citation imbalance. This result supports the use of affirmative action motivated by the goal of reparative justice, rather than simply seeking equality in the moment\cite{Walker2010-WALWIR-6,Olsaretti2018}. The notion of reparative justice foregrounds the importance of restoring historically lost resources, and would support explicit overcitation of women led teams. In contrast, approaches that focus on equality in the moment include distributive approaches, which support the redistribution of existing resources according to some agreed upon moral distribution\cite{Walker2010-WALWIR-6,Olsaretti2018}. Our work suggests that policies incorporating the reparative framework are more likely to effectively overcome imbalances than distributive approaches. 

Even with reparative goals for citing women's scholarship, we show that deeper engagement with scholarship that changes the agent's thinking is needed for sustainable equity. This pattern of findings suggests that programs which support and uplift the work of gender minorities, combined with individual efforts to deeply engage with science beyond one's own network, could be effective methods for mitigating the worsening citation imbalance over time. In neuroscience, organizations tracking and prioritizing the representation of historically excluded groups in talks and conferences such as BiasWatchNeuro (biaswatchneuro.com) and the Innovators in Cognitive Neuroscience seminar series (innovatorsincogneuro.github.io/) exist to counter the dominance of men in speaking opportunities in STEM fields\cite{Teoh2021,Arora2020}.

Being open to learning from new and unfamiliar authors was also a powerful intervention with the potential to sustainably mitigate biased citation practices. Designing and deploying interventions that aim to influence the extent to which authors are open to learning new information would benefit from identifying the psychological mechanisms underlying this behavioral expression of openness. One likely candidate mechanism is intellectual humility\cite{porter2021clarifying}. Intellectual humility is defined by having a non-threatening awareness of one’s intellectual fallibility\cite{krumrei2016development} and being attentive to limitations in the evidence for one’s beliefs and to one’s limitations in obtaining and evaluating relevant information\cite{leary2017cognitive}. Although intellectual humility reflects people’s assessments of their beliefs, it can manifest in an openness to other people’s views and in a flexibility in one’s beliefs and opinions, especially in the light of persuasive evidence. For example, when discussing contentious topics, such as religion, people high in intellectual humility are more tolerant of diverse views\cite{hook2017intellectual}, less willing to perceive their own views as superior\cite{leary2017cognitive}, and more likely to seek knowledge that disavows rather than confirms their views\cite{porter2018intellectual}. Low intellectual humility, by contrast, may manifest in a disregard of viewpoints that differ from one’s own and an unfounded insistence that one’s own beliefs are accurate. Although typically described as a trait or a character strength, there is, as with other related traits (e.g., curiosity\cite{lydon2020within}), evidence for substantial within-person fluctuations in intellectual humility. Indeed, experimental interventions in the form of writing prompts, experiences of awe, or experiences of gratitude successfully increase levels of intellectual humility\cite{kesebir2014quiet,kruse2014upward,stellar2018awe}. As such, we join existing calls for efforts to increase intellectual humility in the scientific process to improve scientific practice\cite{hoekstra2021aspiring}, in this case for the specific goal of mitigating gendered citation biases by promoting an openness to learning from other scholars.

Lastly, we simulated a specific intervention---the citation diversity statement---that we hypothesized had the potential to influence both the total undercitation of women and its change over time. It is likely that the citation diversity statement is a single but important part of the new practices and policies that authors might fruitfully adopt to cite more equitably. Other tools that make transparent authors' perceived gender can be useful during early literature searches (\emph{Citation Transparency} Google Chrome Extension); similarly, tools to identify gender breakdowns in reference, Twitter, and talk compositions can be useful to the field in providing feedback about current gender disparities (see Ref. \cite{Llorens2021} for a list of other resources). Additionally, we recommend a consistent search for openly trans, non-binary, and genderqueer faculty in each author's field, potentially from lists such as 500 queer scientists (500queerscientists.com) and explicit effort to engage with their work, since currently no tools exist for quantifying this engagement despite immense challenges faced by gender nonconforming scientists\cite{Gibney2019}. Lastly, consistent monitoring and data collection on gender and academic outcomes can help hold each field accountable, and identify which programs do and do not work to address disparities.

\subsection*{Limitations and future directions}

We sought to create a model of inequitable citation practices across perceived genders in academia in order to better understand how to cite equitably, though our conclusions should be interpreted in light of the limitations and assumptions of the model. One major limitation is that our name-based algorithmic approach does not allow for a broader analysis of gender-based discrimination against trans, non-binary, and genderqueer scientists whose gender is not probabilistically determinable by their names. These minority genders are often understudied, and so the specific barriers that trans, non-binary, and genderqueer scholars face are currently underquantified \cite{Gillespie2015}, although increasingly qualified \cite{Nicolazzo2016,Pitcher2018}. Existing scholarship confirms that gender minorities broadly construed face substantial structural barriers in academia, similar to other institutions that do not have policies to lower those barriers. The field would greatly benefit from studies of how minority gender identities and expressions affect citation rates in academia.

Additionally, our study purposefully uses a limited range of factors to understand citation practices. This approach allows us to understand minimal, and possibly generalizable factors that lead to population level behaviors, but does not account for other well-documented sources of bias and gender homophily that might impact our results. Future work could expand upon our current model to incorporate biased choices in mentoring, promotion and retention, and awards\cite{Shen2013,Llorens2021}. Similarly, it would be of interest to incorporate the hierarchical social structure of academia, and to better understand the role of non-publishing individuals such as grant managers and journal editors. Lastly, further explorations into which specific types of homophily underlie biases in citations could help target interventions amongst different groups of authors.

Lastly, our work presents one empirical intervention that has the potential to effectively reduce bias in citation practices: the citation diversity statement\cite{Zurn2020,zhou_dale_2020_3672110}. While we demonstrate that authors who are currently using the citation diversity statement show more equitable citations, this data represents the behavior of a relatively small pool of motivated authors, and it is possible that attempts to broadly enforce citation diversity statements would lead to different behavior. We also cannot demonstrate that the use of the citation diversity statement causally influenced citation practices even in this small group. Additionally, while we hypothesize that individuals using the citation diversity statement will be more able to update their internal estimates to keep pace with a diversifying field, we do not yet have empirical evidence to support this hypothesis. Follow-up studies in a few years could investigate the temporal properties of the long-term use of citation diversity statements.

\subsection*{Conclusions}
In this work, we show that observed biases in citation practices can be modeled using only gender homophily in estimates of the field, gender homophily in how those estimates are sampled, and field homophily in meetings. Using these three sources of bias, we are able to recreate an undercitation of authors with woman names, a stronger undercitation from individuals with man names, and an increase in undercitation over time. We show that in our model, sampling field estimates at or above equity (based on field demographics) are not enough to lead to equitable practices. Instead, authors must both sample more women from their network, and become more open to learning from individuals outside their network. Lastly, we set parameters that best recreate the effects of asking authors to use a citation diversity statement in their citations and find that only when a large majority of inequitable citers adopt this policy do we see equitable citations as a field. Our work describes important evidence for homophily as a driver of inequity in science, and provides an understanding of the scope and scale of potential interventions that may serve to alleviate bias.

\section*{Citation Diversity Statement}
Recent work in several fields of science has identified a bias in citation practices such that papers from women and other minority scholars are under-cited relative to the number of such papers in the field \cite{mitchell2013gendered,dion2018gendered,caplar2017quantitative, maliniak2013gender, Dworkin2020.01.03.894378, bertolero2021racial, wang2021gendered, chatterjee2021gender, fulvio2021imbalance}. Here we sought to proactively consider choosing references that reflect the diversity of the field in thought, form of contribution, gender, race, ethnicity, and other factors. First, we obtained the predicted gender of the first and last author of each reference by using databases that store the probability of a first name being carried by a woman \cite{Dworkin2020.01.03.894378,zhou_dale_2020_3672110}. By this measure (and excluding self-citations to the first and last authors of our current paper), our references contain 33.56\% woman(first)/woman(last), 17.07\% man/woman, 19.37\% woman/man, and 30.0\% man/man. This method is limited in that a) names, pronouns, and social media profiles used to construct the databases may not, in every case, be indicative of gender identity and b) it cannot account for intersex, non-binary, or transgender people. Second, we obtained predicted racial/ethnic category of the first and last author of each reference by databases that store the probability of a first and last name being carried by an author of color \cite{ambekar2009name, sood2018predicting}. By this measure (and excluding self-citations), our references contain 8.6\% author of color (first)/author of color(last), 14.65\% white author/author of color, 25.42\% author of color/white author, and 51.32\% white author/white author. This method is limited in that a) names and Florida Voter Data to make the predictions may not be indicative of racial/ethnic identity, and b) it cannot account for Indigenous and mixed-race authors, or those who may face differential biases due to the ambiguous racialization or ethnicization of their names.  We look forward to future work that could help us to better understand how to support equitable practices in science.

\section{Author Contributions}
J.S., K.O, and M.B. analyzed data. J.S. and D.S.B planned analyses. All authors wrote and edited the manuscript.

\section*{Acknowledgements}
We would like to acknowledge the work of the participants of the OHBM Hackathon who helped collect data on individuals who use the citation diversity statement: Isil Poyraz Bilgin; Samuel Guay; Léa Michel; François Paugam; Céline Provins; Adina Wagner.

a

\end{document}